\documentclass[sigconf,screen]{acmart}
\acmBooktitle{Companion Proceedings of the 34th ACM Symposium on the Foundations of Software Engineering (FSE '26), June 5--9, 2026, Montreal, Canada}

\usepackage{balance}
\def\BibTeX{{\rm B\kern-.05em{\sc i\kern-.025em b}\kern-.08em
    T\kern-.1667em\lower.7ex\hbox{E}\kern-.125emX}}

\usepackage{booktabs}
\usepackage{enumitem}
\usepackage{multirow}
\usepackage{subfig}
\usepackage{listings}
\usepackage{color}
\usepackage{threeparttable}
\usepackage{array}
\usepackage{diagbox}
\usepackage{verbatim}
\usepackage{soul}
\soulregister\cite7
\soulregister\ref7
\usepackage{graphicx}
\usepackage{multirow}
\usepackage{hhline}
\usepackage{comment}
\usepackage{float}
\usepackage{array,booktabs}
\usepackage{marvosym} %
\usepackage{pifont}
\usepackage{tcolorbox}
\usepackage{bm}

\usepackage{url}
\usepackage{booktabs}
\usepackage{multirow}
\usepackage{pifont}
\usepackage{graphicx}
\usepackage{textcomp}
\usepackage{listings}
\usepackage{subcaption}
\usepackage{tcolorbox}
\usepackage{xspace}
\usepackage{xcolor}
\usepackage{enumitem}
\usepackage{wrapfig}

\usepackage{algorithmic}
\usepackage[ruled,linesnumbered]{algorithm2e}

\newcommand{\tabincell}[2]{\begin{tabular}{@{}#1@{}}#2\end{tabular}}

\newcommand{\find}[1]{
\begin{tcolorbox}[leftrule=0.5mm,rightrule=0.5mm, toprule=0.5mm,bottomrule=0.5mm,left=2pt,right=2pt,top=2pt,bottom=2pt]%
\em #1
\end{tcolorbox}
}

\newcommand{\ppathf}{{\sc PPatHF}\xspace}

\newcommand{\fixmorph}{{\sc FixMorph}\xspace}

\newcommand{\tsbport}{{\sc TSBPORT}\xspace}

\newcommand{\appname}{{\sc MIP}\xspace}
\newcommand{\appnamebold}{{\sc \textbf{MIP}}\xspace}

\begin{document}
\begin{sloppypar}

\author{Shengyi Pan}
\affiliation{%
\institution{The State Key Laboratory of Blockchain and Data Security, Zhejiang University}
\city{Hangzhou}
\country{China}}
\email{shengyi.pan@zju.edu.cn}

\author{Zhongxin Liu}
\authornote{Corresponding Author}
\authornote{Also with College of Computer Science and Technology, Zhejiang University}
\affiliation{%
  \institution{The State Key Laboratory of Blockchain and Data Security, Zhejiang University}
  \city{Hangzhou}
  \state{Zhejiang}
\country{China}}
\email{liu_zx@zju.edu.cn}

\author{Jiayuan Zhou}
\affiliation{%
\institution{Waterloo Research Center, Huawei}
\city{Waterloo}
\state{Ontario}
\country{Canada}}
\email{jiayuan.zhou1@huawei.com}

\author{Xing Hu}
\affiliation{%
  \institution{The State Key Laboratory of Blockchain and Data Security, Zhejiang University} %
  \city{Ningbo}
\country{China}}
\email{xinghu@zju.edu.cn}

\author{Xin Xia}
\authornote{Also with Hangzhou High-Tech Zone (Binjiang) Institute of Blockchain and Data Security}
\affiliation{%
  \institution{The State Key Laboratory of Blockchain and Data Security, Zhejiang University}
  \city{Hangzhou}
 \country{China}}
\email{xin.xia@acm.org}

\author{Shanping Li}
\affiliation{
  \institution{The State Key Laboratory of Blockchain and Data Security, Zhejiang University}
  \city{Hangzhou}
\country{China}}
\email{shan@zju.edu.cn}

\title{Mitigating Implicit Inconsistencies in Patch Porting}

\begin{abstract}
Promptly porting patches from a source codebase to its variants (e.g., forks and branches) is essential for mitigating propagated defects and vulnerabilities. Recent studies have explored automated patch porting to reduce manual effort and delay, but existing approaches mainly handle inconsistencies visible in a patch’s local context and struggle with those requiring global mapping knowledge between codebases. We refer to such non-local inconsistencies as implicit inconsistencies. 
Implicit inconsistencies pose greater challenges for developers to resolve due to their non-local nature.
To address them, we propose \appname, which enables collaboration among an LLM, a compiler, and code analysis utilities. \appname adopts different strategies for different cases: when source identifiers exist in the target codebase, it leverages compiler diagnostics; otherwise, it retrieves matched code segment pairs from the two codebases as mapping knowledge for mitigation. Experiments on two representative scenarios, cross-fork and cross-branch patch porting, show that \appname successfully resolves more than twice as many patches as the best-performing baseline in both settings. A user study with our industry partner further demonstrates its practical effectiveness.
\end{abstract}

\ccsdesc[500]{Software and its engineering~Software maintenance tools}

\keywords{Patch Porting, Large Language Model, Software Maintenance}

\acmYear{2026}\copyrightyear{2026}
\setcopyright{cc}
\setcctype[4.0]{by}
\acmConference[FSE Companion '26]{34th ACM Joint European Software Engineering Conference and Symposium on the Foundations of Software Engineering}{July 5--9, 2026}{Montreal, QC, Canada}
\acmBooktitle{34th ACM Joint European Software Engineering Conference and Symposium on the Foundations of Software Engineering (FSE Companion '26), July 5--9, 2026, Montreal, QC, Canada}
\acmDOI{10.1145/3803437.3805248}
\acmISBN{979-8-4007-2636-1/26/07}

\maketitle

\vspace{-0.1 cm}
\section{Introduction} \label{sec:introduction}
Forking~\cite{zhou2020thesis, fenske2017variant, zhou2020has} and branching~\cite{zou2019branch, phillips2011branching, shihab2012effect} are widely used to reuse and adapt open source software (OSS) to accommodate different stability and feature requirements. 
In industrial production settings, organizations typically prioritize operational stability and compatibility over rapid adoption of the latest releases, and as a result often deploy older, well-validated versions~\cite{lts_wiki}.
In addition, companies commonly customize OSS to address production-specific needs 
and maintain internal forks to support these variants over long lifecycles~\cite{kessler2009customization,woo2021centris}.
The Linux ecosystem is a representative example of this pattern. 
Enterprise distributions such as OpenEuler~\cite{openeuler} and Red Hat Enterprise Linux~\cite{redhat_linux} are downstream forks that combine vendor-specific customizations with selected stable or long-term support (LTS) branches of the Linux kernel. 

To maintain reliability and security guarantees, downstream variants must continuously port important patches (e.g., security fixes) from their upstream source (e.g., the Linux mainline).
However, the dramatic growth in the number of patches requiring porting poses substantial challenges, as unresponsive porting practices can compromise the stability and security of downstream variants that remain unpatched~\cite{reid2022extent, pan2024automating}.
For example, the Linux kernel saw approximately 3,529 disclosed vulnerabilities in 2024, 
nearly an order-of-magnitude increase over 2023~\cite{linux_vulnerability_surge, linux_vulnerability_surge_2}.
Meanwhile, existing patch porting practices in the kernel ecosystem continue to suffer from significant delays, leaving extended exploit windows open to attackers~\cite{li2024investigation, shariffdeen2020automated}. 
These industrial realities make timely and low-cost patch porting a foundational capability for reducing maintenance burden, shortening security-fix latency, and sustaining production-grade reliability~\cite{lawall2018coccinelle, padioleau2008documenting}.

Many patches cannot be directly ported by copy-and-paste due to the divergence between source and target codebases~\cite{pan2024automating,shariffdeen2021automated}.
For example, Figure~\ref{fig:motivating} presents a patch ported from Vim~\cite{vim_repo} (a widely-used text editor) to Neovim~\cite{neovim_repo} (a fork of vim that focuses on extensibility and usability).
The two code segments share the same functionality and suffer from the same issue.
However, their implementations diverge, e.g., function \verb|dict_find()| is renamed as \verb|tv_dict_find()| in Neovim.
Thus, effective patch porting requires handling implementation differences between codebases~\cite{pan2024automating,shariffdeen2021automated}.

Prior works have explored automated patch porting to reduce the cost and delay of manual porting~\cite{shariffdeen2021automated,pan2024automating,yang2023enhancing}.
Given a patch from the source codebase, existing techniques leverage certain proxies (e.g., syntactic structure~\cite{shariffdeen2021automated} or semantic information~\cite{pan2024automating} of a local code segment) that remain consistent across the two codebases to locate the corresponding code segment in the target codebase and then adapt the patch to local differences.
However, these approaches only address inconsistencies visible in the local context~\cite{shariffdeen2021automated,pan2022automated} and fail to handle inconsistencies that require a global knowledge of differences between codebases (see Section~\ref{subsec:motivating_example} for an example).
We refer to the inconsistencies that are not exposed in the local context as \emph{implicit} inconsistencies.
Implicit inconsistencies commonly arise when patches reference identifiers whose definitions differ across codebases, such as renamed functions or global variables, modified function signatures or variable types.

It is challenging to address the implicit inconsistencies, 
which requires knowledge of the global mappings between the source and the target codebase.
For example, if a function in the source codebase is not defined in the target codebase, one needs to identify how the same functionality is implemented in the target codebase, which may not has a corresponding function but implemented in a completely different way (see Figure~\ref{fig:example_alloc} for an example).
Moreover, such mappings may be consistently updated as the two codebases evolve.
In practice, patch porting is performed exclusively by core project members~\cite{li2024investigation, neovim_top_committer}, who are deeply familiar with both the source and target codebases. 
However, this reliance imposes substantial maintenance burdens and causes significant porting delays~\cite{linux_reduce_LTS, li2024investigation, pan2024automating}. 
For instance, Linux kernel LTS has been reduced from six to two years due to maintenance challenges~\cite{linux_reduce_LTS}.
Investigation with developers at our industry partner also suggests that implicit inconsistencies impose the greatest overhead in patch porting (see Section~\ref{subsec:user_study}).

To this end, we proposes an approach (namely \appname) to automatically address implicit inconsistencies in the ported patches.
\appname is independent of the porting workflow.
It can be used to assist manual patch porting or refine the results of automated porting tools.
Specifically, we first apply the compiler to detect implicit inconsistencies in an incorrect patch, 
as such inconsistencies indicate conflicts with the target project’s context.
For cases where source identifiers exist in the target codebase but have changed attributes, compiler diagnostics provide useful information (e.g., conflicts between identifier usage and definitions). 
We therefore leverage an LLM to resolve these inconsistencies using compiler diagnostics.
In contrast, for the majority cases where source identifiers do not exist in the target codebase, compiler diagnostics alone are insufficient. 
Resolving these non-existent identifiers requires knowledge of how equivalent functionality is implemented across the source and target codebases. 
To acquire this mapping knowledge, we retrieve matched code segment pairs involving the non-existent identifier from both codebases and present them to the LLM as demonstrations to guide the transformation.
Overall, \appname follows an iterative check-and-fix schedule that consists of
\ding{182}~invoking the compiler to check the incorrect patch,
\ding{183}~browsing the source and target codebases to retrieve necessary fixing ingredients,
\ding{184}~invoking the LLM to generate fixes.
This process repeats until no errors remain or a predefined iteration limit is reached.

We conduct evaluations under two typical patch porting scenarios, i.e., cross-fork~\cite{pan2024automating, zhou2020has} and cross-branch~\cite{shariffdeen2021automated, yang2023enhancing}.
Specifically, we follow the existing studies to adopt the patches ported from Vim to Neovim and from the mainline of Linux kernel to older versions for cross-fork and cross-branch evaluation, respectively.
The experiment results verify the effectiveness of \appname in refining ported patches with unresolved implicit inconsistencies.
\appname fixes more than twice as many patches as the best-performing baseline under both scenarios.
Experiment results also verify the generalizability of \appname regarding different underlying LLMs.
Furthermore, we conduct a user study with our industry partner to validate the effectiveness of our prototype and to gather feedback prior to production deployment.
\vspace{-1 mm}
\begin{itemize}[leftmargin=*]
    \item We are the first to propose to mitigate implicit inconsistencies in patch porting, an essential challenge overlooked by the existing studies.
    We design an approach (namely \appname) that facilitates the collaboration among the LLM, compiler, and utilities for analyzing the codebase to mitigate implicit inconsistencies.
    \item We propose to build the mapping knowledge of two codebases by retrieving pairs of matched code segments involving the usage of identifiers, which is effective and offers explainability.
    \item Experiments across two patch porting scenarios show that \appname substantially outperforms baselines, and a user study confirms its practical value and distills lessons learned.
    
\end{itemize} 

\vspace{-0.2cm}
\section{Background and Related Work} \label{sec:relatedwork}
\vspace{-0.1 cm}
In this section, we introduce background of patch porting.

\noindent \textbf{Forking and Branching.}
Forking~\cite{zhou2020thesis, fenske2017variant, zhou2020has} and branching~\cite{zou2019branch, phillips2011branching, shihab2012effect} are widely employed to reuse and customize OSS but inevitability leads to the propagation of 
shared defects among a software family.
Ren~\emph{et al.}~\cite{ren2019automated} conducted an empirical study revealing that 20.5\% of patches require porting across forked projects.
Shariffdeen~\emph{et al.}~\cite{shariffdeen2021automated} find that during 2011 to 2019, over 50k patches made to the mainline of the Linux kernel have been backported to older versions.
Due to the divergence between the source and target codebases, patch porting remains largely a manual task,
which has been identified as inefficient in several studies~\cite{zhou2020has, businge2022reuse, li2024investigation}.
Pan~\emph{et al.}~\cite{pan2024automating} find that over half of the Vim patches being ported to the Neovim take a delay of more than half a year.
Shariffdeen~\emph{et al.}~\cite{shariffdeen2021automated} and Li~\emph{et al.}~\cite{li2024investigation} explore patch porting within the Linux kernel ecosystem, finding that current practices suffer from significant delays.
The delay can be fatal considering security-related flaws~\cite{reid2022extent, pan2024automating}. 
While existing works underscore the importance of patch porting, this paper goes further by addressing implicit inconsistencies, 
a significant challenge in patch porting.

\noindent \textbf{Automated Patch Porting.}
Patch porting aims to transplant a given patch from the source codebase to the target codebase to address a similar issue (e.g., fixing a bug).
Its key challenge is to adapt the patch regarding different implementations of the same functionality between the two codebases~\cite{shariffdeen2021automated, yang2023enhancing, pan2024automating}.
\fixmorph~\cite{shariffdeen2021automated} abstracts the source patch and necessary surrounding context into transformation rules at the syntax level. 
It locates the patch in the target codebase by matching the same pre-patch syntactic structure, while allowing certain generalizations over specific expressions to cope with inconsistencies.
The inconsistencies exposed in the localization process are used to adjust the patch.
\tsbport~\cite{wu2022enhancing} enhances \fixmorph by incorporating program dependency graph matching to align code segments from two codebases, and defines fine-grained patterns to adjust patches.
Recently, \ppathf~\cite{pan2024automating} first leverages LLMs to automate patch porting and achieve substantial improvements over traditional approaches.
Specifically, 
the LLM is prompted to imitate the transformation demonstrated by the pre- and post-patch versions of the source function, and apply it to the pre-patch target function.
With the superiority of LLMs in understanding code semantics, \ppathf can align the code segments from the source and target codebases based on their functionality or semantics, 
and thus is capable of dealing with porting cases where the patch syntactic structure changes greatly.

In summary, existing patch porting techniques leverage certain proxy that remains consistent across the source and target codebase to locate and align a local code unit containing the patch, and adapt the patch according to the inconsistencies exhibited in the context within the local unit.
Unlike existing works, we aim to automate the mitigation of implicit inconsistencies, which is an essential challenge in patch porting while overlooked by existing studies. 
Our approach is orthogonal to existing approaches. 

\section{Motivation and Preliminaries} 
\label{sec:empirical}
\vspace{-0.1cm}

\begin{figure*}[t]
    \centering
    \includegraphics[width=0.9\linewidth]{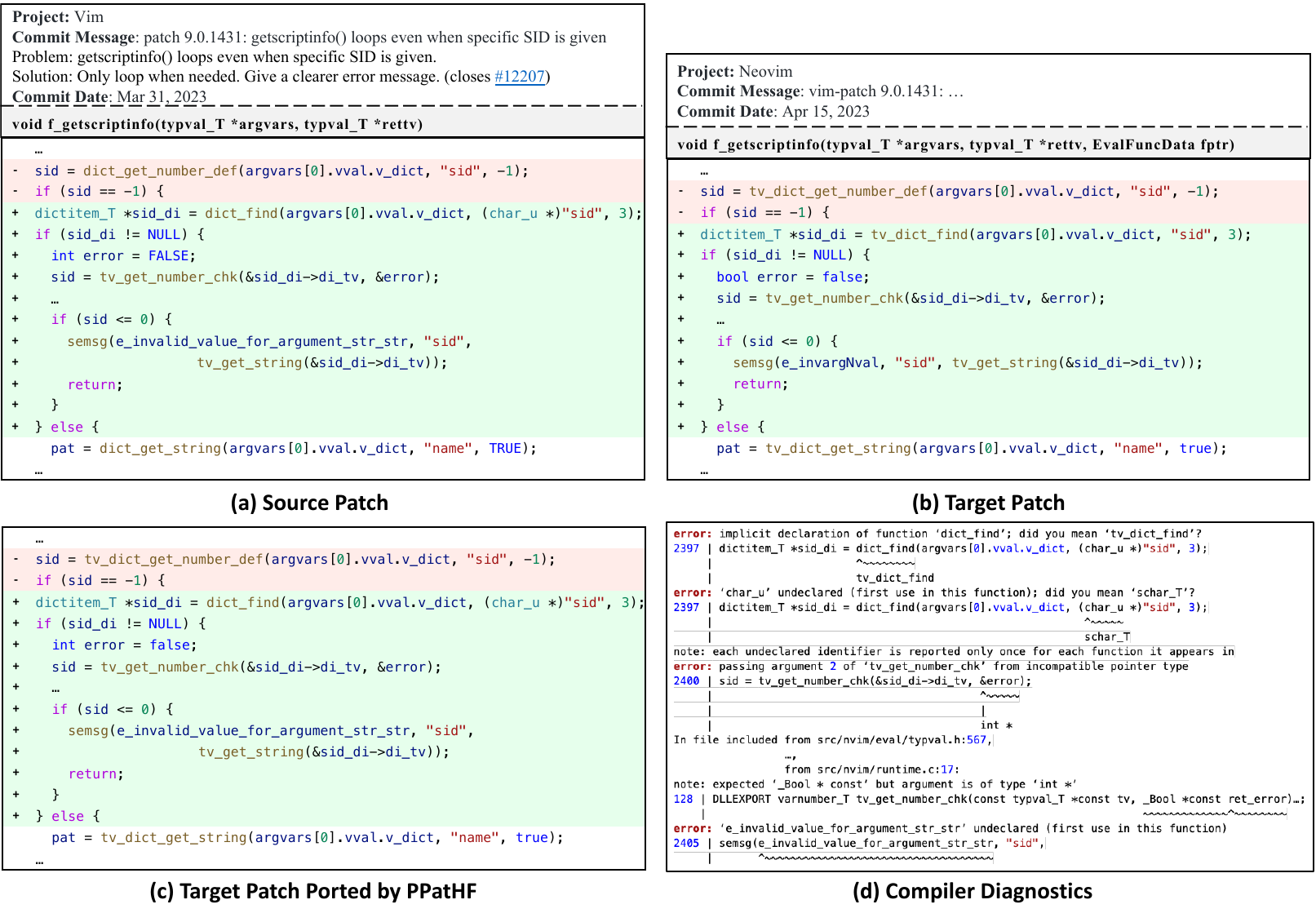}
    \vspace{-0.4 cm}
    \caption{An example of patch porting with implicit inconsistencies}
    \label{fig:motivating}
    \vspace{-0.3 cm}
\end{figure*}

In this section, we provide a motivating example to show the inability of existing patch porting techniques when faced with implicit inconsistencies.
Then, we introduce two specific types of implicit inconsistencies and the challenges to fix them.

\subsection{Motivating Example} \label{subsec:motivating_example}

Figure~\ref{fig:motivating}(a) shows a patch ported from Vim (the source codebase) to Neovim (the target codebase). 
The patch fixes a bug~\cite{vim_12207} in the function \verb|f_getscriptinfo()|, which retrieves script information based on a script ID or a name-based filter.
In the pre-patch version, after extracting the script ID \verb|sid| using \verb|dict_get_number_def()|, the code assumes that an invalid \verb|sid| (i.e., \verb|sid==-1|) implies that the user specified a script name pattern. 
However, an invalid \verb|sid| can also result from supplying a non-positive integer as the script ID.
To fix this bug, the patch refactors the extraction of \verb|sid| into two steps. 
It first attempts to retrieve the item associated with the key \verb|"sid"| from the input dictionary using \verb|dict_find()|. If the item exists (i.e., \verb|sid_di != NULL|), the code then invokes \verb|tv_get_number_chk()| to extract \verb|sid| and reports an error when the value is invalid. 
Because this bug also exists in Neovim due to the initial fork, the patch must be ported to Neovim as well.

As shown in Figure~\ref{fig:motivating}(b), 
porting this patch seems straightforward, as it does not require changes to the patch structure.
However, \ppathf fails to port it correctly, producing an incorrect patch (Figure~\ref{fig:motivating}(c)).
This failure stems from several unresolved inconsistencies between the source and target codebases:
\ding{182}~\verb|dict_find()| is renamed to \verb|tv_dict_find()| in Neovim;
\ding{183}~the second parameter of \verb|dict_find()| changes from \verb|char_u *| to \verb|char *|, making the explicit cast \verb|(char_u *)| unnecessary;
\ding{184}~the second parameter of \verb|tv_get_number_chk()| changes from \verb|int| to \verb|_Bool|, requiring the input argument \verb|error| to be updated from \verb|int| to \verb|bool|; and
\ding{185}~the global constant \verb|e_invalid_value_for_argument_str_str| is renamed to \verb|e_invargNval|.

\ppathf fails to correctly adjust the patch because inconsistencies in identifier definitions (e.g., changes in function signatures or variable names) are implicit: they are not exposed in the modified or context statements of the patch.
In our motivating example, the unresolved inconsistencies involve three identifiers newly introduced by the source patch, i.e., \verb|dict_find()|,  \verb|tv_get_number_chk()|, and \verb|e_invalid_value_for_argument_str_str|, whose definitions are changed in the target codebase.
Moreover, these changes are not reflected in the local context. 
Thus, \ppathf cannot perceive, let alone adapt to, these inconsistencies.
More generally, implicit inconsistencies are difficult to perceive during patch porting because the process typically focuses on applying transformation logic within a localized context, whereas identifying such inconsistencies requires examining patch details in the context of the entire codebase,
i.e., ensuring the definition of every identifier in the patch remains consistent across two codebases.
Addressing implicit inconsistencies is more challenging, even for human developers (see our user study in Section~\ref{subsec:user_study}), as it requires knowledge of global mappings between the source and target codebases
(i.e., how equivalent functionality is implemented in the target codebase).

We aim to propose an approach to automatically address implicit inconsistencies in patch porting.
Specifically, we leverage the compiler as a checker to give feedback on unresolved inconsistencies in ported patches.
Due to inconsistencies in identifier definitions, incorrectly ported patches contradict the context of the target codebase, resulting in compilation errors.
By compiling the target codebase after applying the ported patch, we can accurately identify potential inconsistencies.
For example, Figure~\ref{fig:motivating}(d) shows compiler diagnostics that clearly pinpoint the locations and causes of compilation errors induced by implicit inconsistencies.
In the following sections, we elaborate on addressing implicit inconsistencies based on compiler diagnostics.

\vspace{-0.1cm}
\subsection{Types of Implicit Inconsistencies}
\label{subsec:preliminary_error}
\vspace{-0.1cm}

\begin{figure}
    \centering
    \includegraphics[width=0.8\linewidth]{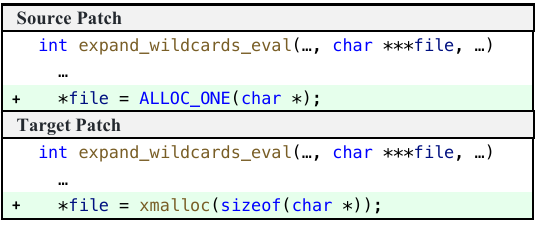}
    \vspace{-0.3 cm}
    \caption{An example of structure changes in patch porting}
    \label{fig:example_alloc}
    \vspace{-0.3 cm}
\end{figure}

We conduct a preliminary study on patches ported from Vim to Neovim (i.e., cross-fork porting) and from the Linux kernel mainline to older versions (i.e., cross-branch porting) to gain a comprehensive view of the types of implicit inconsistencies in patch porting.
We observe that implicit inconsistencies can be classified into two categories based on their causes.

    \noindent \textbf{Type-1: The identifier exists but with changed attributes.}
    This type of inconsistency arises when an identifier exists in the target codebase but its definition has changed, leading to a mismatch between the identifier’s usage and its definition.
    Taking the patch shown in Figure~\ref{fig:motivating} as an example, 
    the type of the second parameter in the function definition of \verb|tv_get_number_chk()| is changed from \verb|int *| to \verb|_Bool *|, resulting in an incompatible pointer type between the input argument and the defined parameter.
    
    \noindent \textbf{Type-2: The identifier does not exist.}
    This type of inconsistency is caused by
    \ding{182}~the namespace conversion of identifiers (e.g., functions, variables, and macros) between the source and target codebases. 
    For example, the global constant \verb|e_invalid_value_for_argument_str_str| is renamed to \verb|e_invargNval| in Figure~\ref{fig:motivating};
    \ding{183}~the identifier involved in the source patch does not have a corresponding one in the target codebase, i.e., the functionality of the identifier is not needed or is implemented in a different way.
For example, in the patch shown in Figure~\ref{fig:example_alloc}, the functionality of the macro \verb|ALLOC_ONE()| in the source patch is implemented through the function \verb|xmalloc()| in the target patch.

\subsection{Challenges to Fix Implicit Inconsistencies}
\label{subsec:preliminary_approach}

We can leverage compiler diagnostics to facilitate the fix of type-1 implicit inconsistencies.
Compiler diagnostics explicitly point out the contradictions between identifier usage and its definition, which can be used to guide the fix.
Taking the patch shown in Figure~\ref{fig:motivating} as an example,
the diagnostics given by the compiler clearly indicates that the second input argument of the function \verb|tv_get_number_chk()| is of type \verb|int *|, while the expected type should be \verb|_Bool * const|.
Additionally, the diagnostics provide the retrieved function definition from the target codebase for reference.

\begin{figure}
    \centering
    \includegraphics[width=0.9\linewidth]{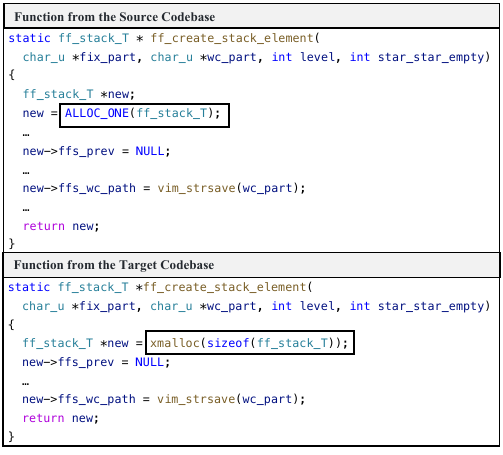}
    \vspace{-0.3 cm}
    \caption{A function pair showing ALLOC\_ONE mapping}
    \label{fig:example_alloc_fix}
    \vspace{-0.5 cm}
\end{figure}

Resolving type-2 inconsistencies (i.e., non-existent identifiers) is more challenging, which requires knowing mapping relationships between the implementations of equivalent functionality in the source and target codebases.
A straightforward approach is to retrieve the code element (e.g., a variable or function) that the identifier points to in the source codebase,
and then search for the most similar element of the same type in the target codebase.
However, this approach has two limitations:
\ding{182}~the content of the identifier can differ significantly across two codebases, leading to incorrect mappings;
\ding{183}~such mappings can only deal with namespace conversions.
However, replacing non-existent identifiers can involve more complex structural transformations, where equivalent functionality is implemented in different ways (see Figure~\ref{fig:example_alloc} for an example).

Our insight is that the identifiers involved in implicit inconsistencies,
typically newly introduced functions, macros, or global variables/constants, are often used in multiple locations across the codebase.
Thus, we can construct code snippet pairs that involve the usage of the identifier between the source and target codebases.
These usage snippet pairs provide an opportunity to infer how the functionality of the identifier is implemented in the target codebase.
Taking the patch shown in Figure~\ref{fig:example_alloc} as an example,
we can retrieve a function from the source codebase that calls \verb|ALLOC_ONE|, as well as its corresponding function in the target codebase (see Figure~\ref{fig:example_alloc_fix}).
This pair of functions share similar implementations, i.e., first create a \verb|ff_stack_T| (a self-defined \verb|Struct|) object and then initialize its attributes using the input parameters.
According to the function pair, we can infer that \verb|ALLOC_ONE()| is used to allocate memory of a given type, and the same functionality in the target codebase is implemented through \verb|xmalloc(sizeof())|.
Such indirect mappings based on the usage context offer several benefits over direct mappings based on the identifier content:
\ding{182}~Since usage snippet pairs present broader context of how the same functionality (involving the identifier usage) is fulfilled across codebases, we can deal with complex mappings that involve structural changes beyond namespace conversions.
\ding{183}~Alleviating mapping errors caused by substantial implementation divergence between codebases.
When conducting mappings based on usage snippets, we can have multiple pairs (i.e., an identifier is typically used multiple times in the codebase) and rely on the promising ones.

\begin{figure}
    \centering
    \includegraphics[width=0.8\linewidth]{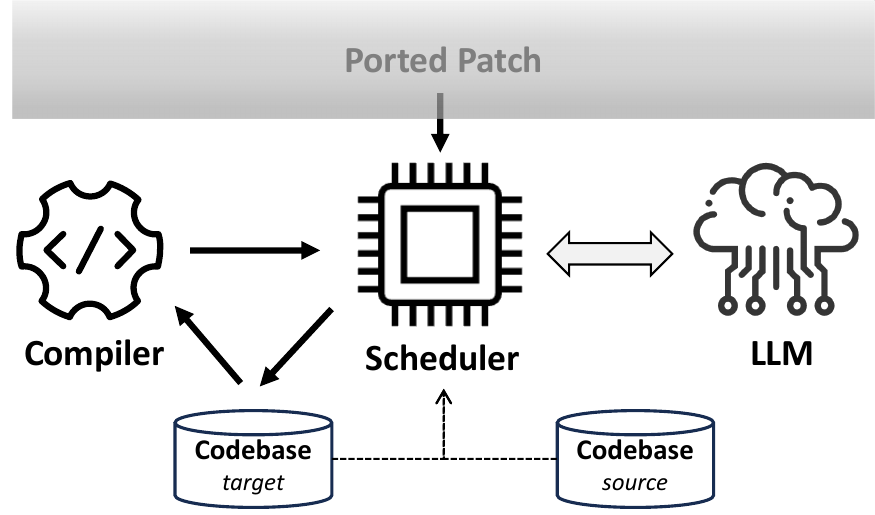}
    \vspace{-0.1 cm}
    \caption{Overview of \appnamebold}
    \label{fig:approach_overview}
    \vspace{-0.3 cm}
\end{figure}

\section{Approach} \label{sec:approach}
In this section, we introduce our proposed approach, namely \appname.
\appname automatically refines ported patches by detecting and addressing implicit inconsistencies. 
We follow the setting of \ppathf~\cite{pan2024automating}, which ports the patch in a function-wise style, to conduct refinement at the function granularity. 
Specifically, given the function applied with the initially ported patch, \appname aims to output a refined version with potential implicit inconsistencies addressed.
Note that \appname is orthogonal to how the patch is initially ported (e.g., manually done by developers or automatically done by existing approaches).
We first give an overview of \appname, then describe two specific mitigation strategies used in its pipeline.

\vspace{-0.1cm}
\subsection{Overview} \label{subsec:overview}
Figure~\ref{fig:approach_overview} presents the overview of \appname, consisting of three modules:
\begin{itemize}[leftmargin=*]
\item \textbf{Scheduler} executes our designed mitigation plan. 
It first invokes the compiler to perform compilation and parses compiler diagnostics.
Next, it browses the source and target codebases to retrieve fixing ingredients based on these diagnostics.
Specifically, given an identifier, Scheduler can retrieve its definition and code snippets that involve its usage.
Finallly, Scheduler invokes the LLM to execute instructions and parses the responses.

\item \textbf{LLM} performs inferences (e.g., fixing specified errors) as instructed by the Scheduler.

\item \textbf{Compiler} compiles the target codebase and outputs diagnostics.
As discussed in Section~\ref{sec:empirical}, implicit inconsistencies lead to compilation errors, and the compiler diagnostics offer information that facilitates error fix, including the exact location, detailed causes, and potential fixing suggestions.
\end{itemize}
\vspace{-0.1cm}

\begin{algorithm}[t]
\small
\caption{Refinement of Ported Patch }
\label{alg:overview}
\SetKwData{functarget}{tgtFunc}  %
\SetKwData{functargetrefined}{tgtFuncRefin}  %
\SetKwData{codebasetarget}{tgtCB}
\SetKwData{codebasesource}{srcCB}
\SetKwData{maxiter}{maxIter}
\SetKwData{currentiter}{curIter}
\SetKwData{errlist}{errList}
\SetKwData{errlistnoidentifier}{noIDErrList}
\SetKwData{errnoidentifier}{noIDErr}
\SetKwProg{Fn}{Function}{:}{}
\SetKwFunction{applypatch}{apply\_patch}
\SetKwFunction{compile}{compile}
\SetKwFunction{filtertnonexistidentifiererr}{get\_nonexist\_identifier\_err}
\SetKwFunction{fixbyreferencecontext}{fix\_by\_usage\_pairs}
\SetKwFunction{fixbydiagnostics}{fix\_by\_diagnostics}
\SetKwInOut{Input}{Input}
\SetKwInOut{Output}{Output}
\SetKw{Break}{break}

\Fn{\appname}{
    \Input{\functarget (target function applied with initially ported patch), \codebasetarget (target codebase), \codebasesource (source codebase), \maxiter (max iterations)}
    \Output{\functargetrefined (target function with implicit inconsistencies refined)}
    \BlankLine
    \functargetrefined, \currentiter $\leftarrow$ \functarget, 0\\
    \codebasetarget $\leftarrow$ \applypatch(\codebasetarget, \functargetrefined)\\ \label{alg_overview:update}
    \errlist $\leftarrow$ \compile(\codebasetarget)\\ \label{alg_overview:compile}
    \While{\errlist $\ne \emptyset$ and \currentiter $<$ \maxiter}{ \label{alg_overview:stop}
        \errlistnoidentifier $\leftarrow$ \filtertnonexistidentifiererr(\errlist) \\ \label{alg_overview:select_namespace_err}
        \uIf{\errlistnoidentifier is not NONE}{ 
            \For{\errnoidentifier in \errlistnoidentifier}{
                \functargetrefined $\leftarrow$ \fixbyreferencecontext(\functargetrefined, \errnoidentifier, \codebasesource, \codebasetarget) \label{alg_overview:fix_by_reference_context}
            }
        }
        \Else{
            \functargetrefined $\leftarrow$ \fixbydiagnostics(\functargetrefined, \errlist, \codebasetarget) \label{alg_overview:fix_by_diagnostics}
        }
        \currentiter $\leftarrow$ \currentiter$+1$\\
        \codebasetarget $\leftarrow$ \applypatch(\codebasetarget, \functargetrefined)\\ \label{alg_overview:update1}
        \errlist $\leftarrow$ \compile(\codebasetarget)\\ \label{alg_overview:compile1}
    }
    \algorithmicreturn{ \functargetrefined }
}
\end{algorithm}

Algorithm~\ref{alg:overview} details our proposed patch refinement process for mitigating implicit inconsistencies.
To start with, 
given the function applied with the initially ported patch,
\appname first uses it to replace the corresponding unpatched function in the target codebase (Line~\ref{alg_overview:update}).
Then, \appname runs the compiler to compile the updated codebase and retrieves the initial compiler diagnostics (Line~\ref{alg_overview:compile}).
Based on the initial diagnostics, \appname iteratively refines the patched function.
The refinement process iterates until either no compilation errors are detected, implying that all inconsistencies have been resolved, or a pre-set maximum iteration number is reached (Line~\ref{alg_overview:stop}).
In each iteration, \appname first fixes certain errors (Lines~\ref{alg_overview:select_namespace_err}-\ref{alg_overview:fix_by_diagnostics}) and then refresh the diagnostics (Line~\ref{alg_overview:update1}\&\ref{alg_overview:compile1}).

Regarding the specific refinement process in each iteration (Lines~\ref{alg_overview:select_namespace_err}-\ref{alg_overview:fix_by_diagnostics}),
\appname first checks for errors caused by non-existent identifiers (Line~\ref{alg_overview:select_namespace_err}).
\appname prioritizes fixing this type of error, as is can be the cause of other errors.
A typical example is the implicit function declaration error, i.e., a function is called before declaration~\cite{gcc_manual}.
The C compiler proceeds with compilation by generating a default implicit declaration, specifying an integer return type and no parameters. 
Typically, this implicit declaration does not match the intended function signature, leading to associated errors.
\appname fixes non-existent identifier errors one by one based on mapping knowledge derived from retrieved usage code snippet pairs from two codebases, as discussed in Section~\ref{subsec:preliminary_approach}.
The details of \verb|fix_by_usage_pairs()| are introduced in Section~\ref{subsec:fix_by_reference_context}.
Once finished mitigating these errors, \appname directly transitions to the next refinement iteration, skipping any remaining errors in the current iteration and refreshing the diagnostics.
If no non-existent identifier error is detected in the current iteration, \appname simultaneously fixes all remaining errors based on compiler diagnostics (see Section~\ref{subsec:fix_by_compiler_message}).

\begin{figure}[t]
    \centering
    \includegraphics[width=\linewidth]{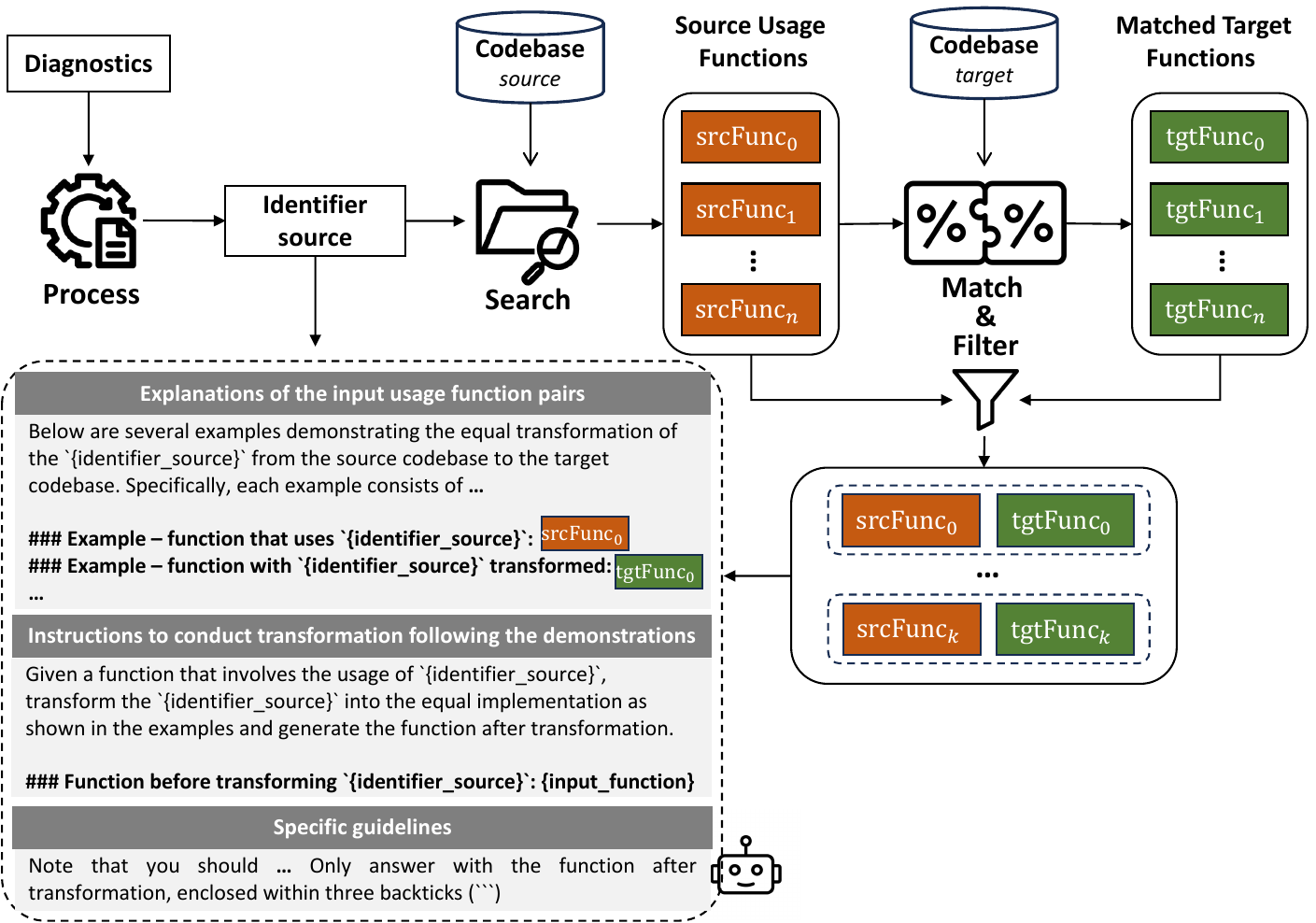}
    \vspace{-0.1 cm}
    \caption{Pipeline of fixing non-existent identifiers through matched pairs of usage code snippets}
    \vspace{-0.3 cm}
    \label{fig:pipeline_fix_by_reference_context}
\end{figure}

\subsection{Fix by Matching Pairs of Usage Snippets} \label{subsec:fix_by_reference_context}
As discussed in Section~\ref{subsec:preliminary_approach}, \appname fixes non-existent identifier errors by leveraging mapping knowledge derived from matched usage function pairs
across two codebases.
Figure~\ref{fig:pipeline_fix_by_reference_context} shows the pipeline.
We first extract the source identifier, i.e., the non-existent identifier to be replaced in the ported patch, from the compiler diagnostics.
Next, we search the source codebase to retrieve
code snippets (functions in our implementation) 
that use this identifier.

For each retrieved source function, 
we try to identify a corresponding function in the target codebase based on text and semantic similarities.
We first narrow down the pool of candidates by selecting the top $n$ (i.e., five in our implementation) functions with the highest text similarities.
Specifically, after preprocessing (e.g., removing comments and parsing code into tokens), we consider a hierarchy of text similarities at three levels of granularity:
the function name, declaration, and body.
We adopt this hierarchical approach, rather than directly comparing entire functions, to place greater emphasis on similarities in function declarations, which are generally more stable and less prone to change than function bodies.
All functions from the target codebase are ranked by the sum of three similarity measures.
For the top $n$ candidates, we further utilize the LLM to identify the one that is semantically equivalent to the given source function, thereby reducing potential mapping errors of relying solely on text similarity.

Furthermore, since a global identifier is typically used multiple times throughout the codebase, we can retrieve multiple usage function pairs for it.
To ensure the quality of the demonstrations used to guide the LLM, we further introduce two strategies to filter the candidate usage function pairs.
First, we exclude usage function pairs with name or body similarity falls below preset thresholds.
We seek to present the LLM with function pairs that are as similar as possible, since implementation differences could confuse the LLM in extracting the correct transformations of the designated identifier.
Second, we sort the usage function pairs to select the top $k$ (i.e., three in our implementation) pairs with the shortest lengths.
We aim to present the LLM with functions as concisely as possible, as excessive irrelevant context could distract the LLM from focusing on the transformation of the designated identifier.

After retrieving the candidate usage function pairs, we use them to build the prompt as shown in Figure~\ref{fig:pipeline_fix_by_reference_context}.
The prompt is parameterized over error-specific content, indicated with the `$\{\}$' syntax.
The prompt consists of three sections.
The first section instructs the LLM on how to interpret input usage function pairs, i.e., examples showing the transformations of the identifier from the source codebase to the target codebase.
The second section directs the LLM to replace the use of the source identifier in the function to be refined, 
following the transformations shown in the provided usage function pairs.
The third section offers specific guidelines to align the LLM's behavior with our expectations (e.g., output format).

\vspace{-0.1cm}
\subsection{Fix by Compiler Diagnostics} \label{subsec:fix_by_compiler_message}
\begin{figure}
    \centering
    \includegraphics[width=0.85\linewidth]{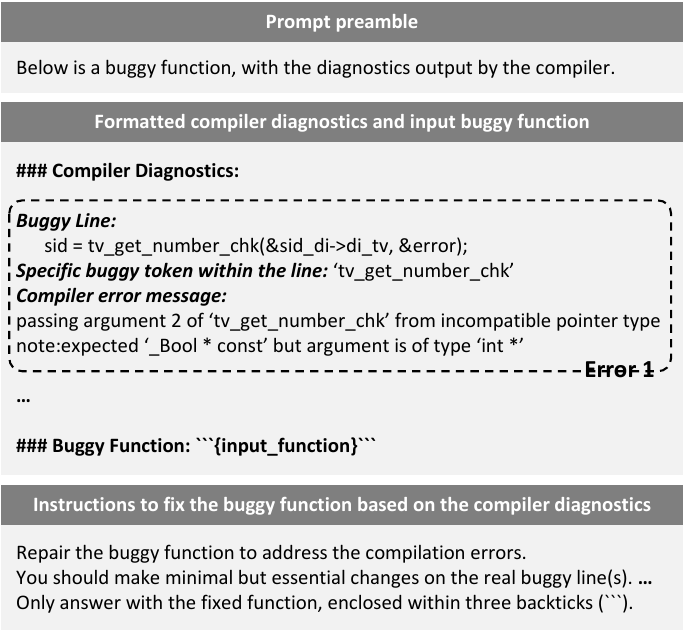}
    \vspace{-0.1 cm}
    \caption{Prompt template for fixing by compiler diagnostics}
    \label{fig:prompt_fix_by_compiler_diagnostics}
    \vspace{-0.3 cm}
\end{figure}

For errors caused by changed definitions of the same identifier, we instruct the LLM to simultaneously fix all of them based on compiler diagnostics.
Figure~\ref{fig:prompt_fix_by_compiler_diagnostics} presents the prompt constructed for our motivating example (see Figure~\ref{fig:motivating}).
The prompt consists of three sections:
a preamble, error-wise formatted diagnostics alongside the input buggy function, and instructions to guide the LLM to perform fix according to the compiler diagnostics.
Rather than employing the infilling-style prompt commonly adopted by existing program repair works~\cite{xia2022less, joshi2023repair, wang2023rap} (i.e., asks the LLM to fill in a specific masked buggy line), we use a refinement-style prompt (i.e., ask the LLM to refine the entire function).
The rationale for this design is that a fix may not occur at the specific buggy location identified by the compiler, but in relevant contexts instead. 
For example, to address the incompatible pointer type of the input argument \verb|error| against the definition of \verb|tv_get_number_chk()| shown in Figure~\ref{fig:motivating}, one needs to modify the variable type in the definition of \verb|error|.

The diagnostics for each compilation error consist of two parts: the error location and the error message, which may be accompanied by additional notes providing necessary context for interpreting the message.
Instead of directly supplying the LLM with the text format of compiler diagnostics, we set the compiler to output the raw structured diagnostics (e.g., in \verb|json| format) and convert it into an LLM-friendly format.
Specifically, the text format of diagnostics visualizes the error with markers (see Figure~\ref{fig:motivating}(d)) to allow human developers to intuitively understand it.
However, such visualization can be confusing for the LLM as it interprets the input as a sequence of tokens. 
We format the diagnostics in a straightforward way for the LLM to interpret, including three segments:
1) the specific buggy line extracted from the function,
2) the specific buggy token within the buggy line,
3) the error message, together with supplementary explanations.

\section{Evaluation} \label{sec:evaluation}
In this section, we first introduce our experiment setting. Then, we describe the experiment results. 

\subsection{Experiment Setting}\label{subsec:experiment_setting}
We aim to answer the following two RQs:
\begin{itemize}[leftmargin=*]
    \item \textbf{RQ1: How effective is \appnamebold in mitigating implicit inconsistencies during patch porting?}
    \item \textbf{RQ2: How effective is \appnamebold with different LLMs?}
\end{itemize}

\noindent \textbf{Dataset.}
We consider two typical patch porting scenarios in our experiments: cross-fork~\cite{zhou2020has, pan2024automating} and cross-branch~\cite{shariffdeen2021automated, yang2023enhancing}.
For cross-fork porting, we use the dataset from \ppathf~\cite{pan2024automating}, which contains patches ported from Vim to Neovim.
For cross-branch porting, we use the datasets from \fixmorph~\cite{shariffdeen2021automated} and \tsbport~\cite{yang2023enhancing}, which contain patches backported from Linux mainline to older versions.
Since this work focuses on automatically mitigating implicit inconsistencies in ported patches, 
we further collect incorrectly ported patches with unresolved inconsistencies for evaluation.
To avoid the high cost of manual porting,
we apply state-of-the-art techniques (i.e., \ppathf for Vim-Neovim and \tsbport for Linux) to port patches automatically and collect those that trigger compilation errors.
Because automatically ported patches may also contain other types of errors (e.g., missing patch logic or incorrect patch locations) that are beyond the scope of this work, we manually curate the results to retain only errors caused by implicit inconsistencies.
Our user study (see Section~\ref{subsec:user_study}) shows that implicit inconsistencies are more challenging for developers during manual validation, whereas other porting errors can usually be resolved using local information.

Finally, the Vim-Neovim dataset contains 64 patch porting samples with 143 implicit inconsistencies; one sample may contain multiple inconsistencies.
Among them, 37 (25.8\%) are type-1 and 106 (74.1\%) are type-2.
The Linux backporting dataset contains 44 samples with 62 implicit inconsistencies.
All of them are type-2, possibly because the source and target codebases are less divergent in cross-branch porting~\cite{pan2024automating}.

\noindent \textbf{Validation Process.}
Following existing patch porting studies~\cite{shariffdeen2020automated, shariffdeen2021automated, yang2023enhancing}, we measure effectiveness by the number (percentage) of \textit{correct} fixes, where a correct fix is defined as one that is semantically equivalent to the ground truth.
Since a ported patch may contain multiple implicit inconsistencies, we report results at both the patch and inconsistency levels for a more comprehensive evaluation.
We manually validate each fix, following prior work on automated patch porting~\cite{shariffdeen2021automated, yang2023enhancing}.
Specifically, two authors independently perform the following steps:
\ding{182}~For each implicit inconsistency, we check the compiler diagnostics after applying the refined patch to ensure that the original compilation errors are resolved and no new compilation errors are introduced.
\ding{183}~We verify the semantic equivalence between the fix and the ground truth.
\ding{184}~If all implicit inconsistencies in a patch are correctly fixed without introducing irrelevant modifications, we regard it as a correct patch-level fix.
After the independent annotations, the two authors discuss and resolve disagreements.
The Cohen's Kappa coefficient between the two annotators is 0.82, indicating strong inter-annotator agreement.

\noindent \textbf{Implementation Details.} 
We implement \appname with Llama3.1-70b~\cite{dubey2024llama} as the underlying LLM.
We use gcc~\cite{gcc_manual} as the proxy of the \textit{Compiler} in \appname, and set the output format of compiler diagnostics to \verb|json|.
The functionality of \appname that retrieves the definition and references of a given identifier from the codebase
is implemented using GNU Global (i.e., a source code tagging system~\cite{gnu_global}).
We implement the source code analysis functionality of \appname (e.g., function signature extraction) using Tree-sitter~\cite{treesitter}.

\subsection{RQ1. The Effectiveness of \appnamebold}~\label{subsec:RQ1}

\noindent \textbf{Method.} 
We evaluate the effectiveness of \appname in mitigating implicit inconsistencies on the test sets collected in Section~\ref{subsec:experiment_setting}.
As the first study on mitigating implicit inconsistencies in patch porting, there is no direct baseline.
We therefore include the following baselines:
\ding{182}~\emph{Compiler Suggestion (CS).}
For type-2 implicit inconsistencies (i.e., non-existent identifiers), compiler diagnostics may provide hints for potential fixes.
For example, as shown in Figure~\ref{fig:motivating}(d), the compiler suggests replacing \verb|dict_find| with \verb|tv_dict_find|.
Accordingly, this baseline replaces non-existent identifiers with compiler-recommended ones.
\ding{183}~\emph{Compiler Suggestion Enhanced by Function Definition Matching (CS+).}
For non-existent function calls, a more precise fix may be obtained by matching function definitions rather than relying only on compiler suggestions.
Specifically, for each non-existent function detected by the compiler, we first retrieve its definition from the source codebase and then search the target codebase for the function with the most similar definition.
We adopt the same similarity measure as \appname (see Section~\ref{subsec:fix_by_reference_context}).
\ding{184}~\emph{Llama3.1-70b}~\cite{dubey2024llama}.
We also include the underlying LLM used in \appname as a baseline.
Specifically, we keep all settings the same as \appname, except that for type-2 inconsistencies, we provide Llama with compiler diagnostics directly instead of retrieved pairs of usage snippets.
Comparing \appname with Llama helps evaluate the effectiveness of our design, especially the use of usage-snippet pairs to guide the mitigation of type-2 inconsistencies.

\begin{table}[tbp]
  \centering
  \caption{Performance comparisons between \appnamebold and baselines for mitigating implicit inconsistencies}
  \vspace{-0.3cm}
  \scalebox{0.8}{    
    \begin{tabular}{clcccccc}
    \toprule
    \multirow{2}[4]{*}{\textbf{Task}} & \multicolumn{1}{c}{\multirow{2}[4]{*}{\textbf{Method}}} & \multicolumn{2}{c}{\textbf{Patch Level}} & \multicolumn{4}{c}{\textbf{Inconsistency Level}} \\
\cmidrule{3-8}          &       & \textbf{\#Patch} & \textbf{Patch\%} & \textbf{\#T1} & \textbf{T1\%} & \textbf{\#T2} & \textbf{T2\%} \\
    \midrule
    \multirow{4}[2]{*}{\tabincell{c}{Vim\\Neovim}} & CS    & 2 / 64 & 3.1\% & 0 / 37 & 0.0\% & 9 / 106 & 8.5\% \\
          & CS+   & 5 / 64  & 7.8\% & 0 / 37 & 0.0\% & 15 / 106 & 14.2\% \\
          & Llama & 26 / 64  & 40.6\% & 27 / 37 & 73.0\% & 55 / 106 & 51.9\% \\
          & \appname & \textbf{52 / 64} & \textbf{81.3\%} & \textbf{33 / 37} & \textbf{89.2\%} & \textbf{98 / 106} & \textbf{92.5\%} \\
    \midrule
    \multirow{4}[2]{*}{\tabincell{c}{Linux\\Backport}} & CS    & 0 / 44 & 0.0\% & /     & /     & 1 / 62 & 1.6\% \\
          & CS+   & 4 / 44  & 9.1\% & /     & /     & 5 / 62 & 8.1\% \\
          & Llama & 13 / 44  & 29.5\% & /     & /     & 21 / 62 & 33.9\% \\
          & \appname & \textbf{32 / 44} & \textbf{72.7\%} & /     & /     & \textbf{50 / 62} & \textbf{80.6\%} \\
    \bottomrule
    \end{tabular}}
    \vspace{-0.3cm}
  \label{table:RQ1}%
\end{table}%

\smallskip
\noindent \textbf{Results.}
Table~\ref{table:RQ1} compares \appname with the baselines in mitigating implicit inconsistencies during patch porting, where T1 and T2 denote type-1 and type-2 inconsistencies, respectively.
For the \emph{CS} baseline, it fixes only a very limited number (less than 10\% on both datasets) of type-2 inconsistencies.
Based on our manual review, compiler suggestions for non-existent identifiers are mainly heuristic, typically searching for identifiers of the same type with the most similar names.
As a result, they are effective only for simple namespace conversions (i.e., without structural changes) and when the corresponding identifier in the target codebase has a similar name.
For \emph{CS+}, incorporating definition-based matching brings only limited improvement. 
Compared with \emph{CS}, it correctly fixes six more type-2 inconsistencies (from 9 to 15) on Vim-Neovim and four more (from 1 to 5) on Linux.
We identify two main reasons for its limited effectiveness.
First, like compiler suggestions, it cannot handle fixes involving structural changes.
Second, definition-based matching may fail when the source and corresponding target function implementations have significantly diverged.
The poor performance of \emph{CS} and \emph{CS+} supports our motivation for inferring indirect mappings from identifier usage context, rather than relying on direct mappings based on identifier content.

Regarding \appname, it correctly fixes 81.3\% of patches on Vim-Neovim and 72.7\% on Linux.
Compared with Llama, \appname improves patch-level performance by over 100\% on both datasets, mainly due to its stronger ability to handle type-2 inconsistencies.
Specifically, \appname correctly fixes 78.2\% and 138.1\% more type-2 inconsistencies on the two datasets, respectively.
This improvement demonstrates the effectiveness of retrieving pairs of code segments involving the non-existent identifier to inject mapping knowledge into the LLM.
Essentially, the \emph{off-the-shelf} Llama lacks mapping knowledge between the source and target codebases.
Instead, it mainly \emph{recalls} candidate implementations from the given context, relying on common code patterns memorized during pretraining.
After manually inspecting Llama's results, we find that most correctly mitigated type-2 inconsistencies fall into two categories:
1) simple namespace conversions, and
2) common global constants or utility functions widely used in the target codebase, e.g., replacing \verb|vim_free()| with \verb|xfree()| for memory deallocation.
However, this \emph{recall}-based strategy lacks rationale and is unreliable.
\ding{182}~Its effectiveness is highly context-dependent; for example, we observe that the same non-existent identifier may be fixed differently across contexts, such as incorrectly fixing \verb|vim_free()| by adding a type cast to its parameter.
\ding{183}~Since both source and target codebases continuously evolve, memorized knowledge from pretraining can also become outdated.
This further highlights an advantage of \appname. Specifically, unlike the black-box \emph{recall} of LLMs, our approach is evidence-based, with fixes guided by retrieved pairs of usage functions that explicitly show how the source identifier should be transformed.
These usage function pairs also make it easier for developers to validate and adjust the suggested fixes.

In addition, \appname increases the number of correctly fixed type-1 inconsistencies by 22.2\% on Vim-Neovim.
This is because providing the LLM with all compilation errors at once can be confusing, especially when some errors are caused by non-existent identifiers (see Section~\ref{subsec:overview}).
To address this, \appname adopts a schedule that first resolves non-existent identifiers and refreshes compiler diagnostics before handling other errors.

\begin{figure}[t]
  \centering
  \includegraphics[width=0.8\linewidth]{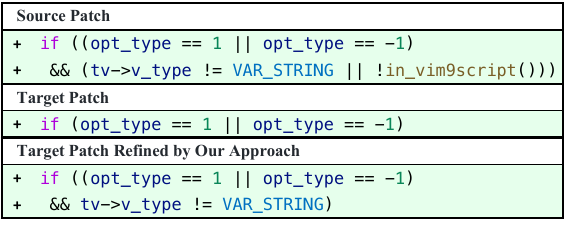}
    \vspace{-0.1 cm}
    \caption{An example of failed mitigation}
    \vspace{-0.3 cm}
    \label{fig:example_failed_invim9script}
\end{figure}

\noindent \textbf{Failure Cases.}
We manually analyze the failure cases of \appname and categorize them into two main types:
\ding{182}~No valid code-segment pairs involving the identifier usage can be retrieved, or the retrieved pairs do not demonstrate valid and generalizable transformations.
Such failures usually involve identifiers that are rarely used in the codebases.
\ding{183}~\appname (specifically, its underlying LLM) fails to accurately reason about the transformations shown in the retrieved code-segment pairs.
These failures typically involve complex transformations.
For example, for \verb|in_vim9script()| (see Figure~\ref{fig:example_failed_invim9script}), \appname successfully removes \verb|in_vim9script()| in the refined patch, as indicated by the retrieved pairs, but fails to also remove the associated condition (i.e., \verb|tv->v_type != VAR_STRING|) that depends on it.

\find{
{\bf RQ1:}
Matching the corresponding identifier in the target codebase using only the identifier’s own information is ineffective.
\appname outperforms the baselines significantly.
}

\subsection{RQ2. The Generalizability of \appnamebold}~\label{subsec:RQ2}

\noindent \textbf{Method.}
In RQ2, we evaluate the generalizability of \appname across different underlying LLMs.
Specifically, we replace the default LLM with two additional models
(i.e., GPT-4o~\cite{gpt-4o} and DeepSeek-v3~\cite{liu2024deepseek})
while keeping all other settings unchanged.
We conduct this evaluation on the Vim-Neovim dataset, which exhibits a larger gap between the source and target codebases.

\begin{figure}[t]
  \centering
  \includegraphics[width=0.5\linewidth]{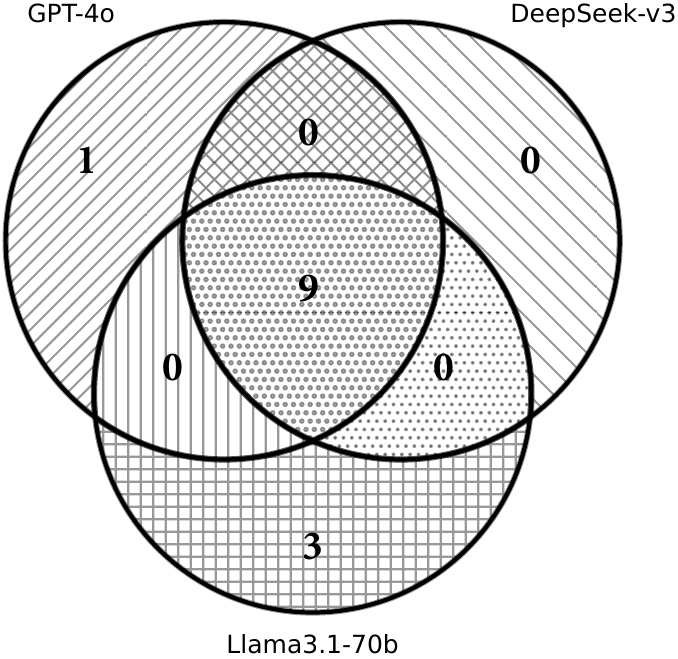}
    \vspace{-0.1 cm}
    \caption{Venn diagram of failure cases of different LLMs}
    \vspace{-0.3 cm}
    \label{fig:venn_failed_cases}
\end{figure}

\noindent \textbf{Results.}
Table~\ref{table:RQ2} shows the performance of \appname with different underlying LLMs.
Figure~\ref{fig:venn_failed_cases} further shows the overlap of failure cases among the three models.
\appname-\emph{DeepSeek} and \appname-\emph{GPT} slightly outperform \appname-\emph{Llama},
likely due to differences in the underlying coding abilities of the LLMs, as DeepSeek-v3 and GPT-4o have reported stronger performance than Llama3.1-70b on general coding benchmarks~\cite{liu2023is, evalplus_leaderboard}.
Notably, Figure~\ref{fig:venn_failed_cases} shows that the three variants share a large portion of failure cases.
Overall, these results suggest that \appname works effectively and consistently across different underlying LLMs, although its performance is still influenced by their coding capabilities.

\begin{table}[tbp]
  \centering
  \caption{Performance of \appnamebold with different underlying LLMs}
  \vspace{-0.3cm}
  \scalebox{0.85}{
    \begin{tabular}{lcccccc}
    \toprule
    \multicolumn{1}{c}{\multirow{2}[4]{*}{\textbf{Approach}}} & \multicolumn{2}{c}{\textbf{Patch Level}} & \multicolumn{4}{c}{\textbf{Inconsistency Level}} \\
\cmidrule{2-7}          & \textbf{\#Patch} & \textbf{Patch\%} & \textbf{\#T1} & \textbf{T1\%} & \textbf{\#T2} & \textbf{T2\%} \\
    \midrule
    \appname-\emph{Llama} & 52 / 64 & 81.3\% & 33 / 37 & 89.2\% & 98 / 106 & 92.5\% \\
    \appname-\emph{GPT} & 54 / 64 & 84.4\% & \textbf{34 / 37} & \textbf{91.9\%} & 98 / 106 & 92.5\% \\
    \appname-\emph{DeepSeek} & \textbf{55 / 64}  & \textbf{85.9\%} & \textbf{34 / 37} & \textbf{91.9\%} & \textbf{99 / 106} & \textbf{93.4\%} \\
    \bottomrule
    \end{tabular}}
    \vspace{-0.1cm}
  \label{table:RQ2}%
\end{table}

\find{
{\bf RQ2:}
\appname performs consistently across different underlying LLMs, with better performance achieved by stronger coding models.
}

\section{Discussion} \label{sec:discussion}
In this section, we discuss the practical application of \appname, the lessons learned and the threats to validity.

\subsection{User Study} \label{subsec:user_study}
We conduct a user study with developers at our industry partner to validate the effectiveness of our prototype.

\noindent \textbf{Design.}
We select ten patch porting cases from Vim-Neovim, all involving implicit inconsistencies.
For each case, participants are provided with
a)~the source patch and source codebase,
b)~the target codebase to be patched, and
c)~the patch initially ported by \ppathf~\cite{pan2024automating} and further refined by \appname, which serves as a reference.
Participants are asked to port the source patch to the target codebase and submit the patched version.
During the process, they can run \appname to check the ported patch for unresolved implicit inconsistencies and obtain suggested fixes.
We provide the compiler diagnostics, the retrieved usage function pairs referenced by \appname, and the generated fix.

We ensure that the selected cases do not overlap in the specific implicit inconsistencies involved.
Among them, seven cases are ones where \appname correctly fixes the implicit inconsistencies, and three are ones where it fails.
We include the failed cases to provide a more comprehensive view of \appname's impact on patch porting.
We recruit six software engineers from our industry partner as participants.
The six participants are randomly divided into two groups of three:
a)~the experimental group uses \appname, which provides diagnostics, suggested fixes, and retrieved usage function pairs;
b)~the control group uses only the compiler, which provides diagnostics alone.

Before the study, we introduce the patch porting task and explain how to use \appname.
To ensure familiarity with the task and tool, all participants complete three pilot tasks before the formal study.
Each task has a time limit of 10 minutes, and participants may submit as soon as they finish.
We record task completion time and participants' ratings of \appname's usefulness on a 5-point Likert scale.
If a participant fail to submit within the time limit, we mark the task as failed and ask them to explain the obstacles preventing completion.
We also record the modifications participants make each time they run \appname/compiler for post-experiment analysis.

\begin{table}[tbp]
  \centering
  \caption{User study results}
  \vspace{-0.1cm}
  \scalebox{0.68}{
    \begin{tabular}{|l|c|c|c|c|c|c|c|c|c|c|c|}
    \hline
    \multicolumn{2}{|c|}{\multirow{2}[4]{*}{Tasks}} & \multicolumn{7}{c|}{Correct Cases}                    & \multicolumn{3}{c|}{Failed Cases} \\
\cline{3-12}    \multicolumn{2}{|c|}{} & T1    & T2    & T3    & T5    & T6    & T7    & T9    & T4    & T8    & T10 \\
    \hline
    \multicolumn{1}{|c|}{\multirow{2}[4]{*}{\#Correct}} & Exp.  & 3     & 3     & 3     & 3     & 3     & 3     & 3     & 0     & 0     & 0 \\
\cline{2-12}          & Ctrl. & 0     & 3     & 3     & 1     & 3     & 3     & 1     & 0     & 0     & 0 \\
    \hline
    \multicolumn{1}{|c|}{\multirow{2}[4]{*}{Time(s)}} & Exp.  & 378.7  & 164.3  & 138.3  & 256.0  & 369.0  & 166.3  & 369.7  & 333.0  & /     & 269.7  \\
\cline{2-12}          & Ctrl. & 539.5  & 213.0  & 296.3  & 409.7  & 547.3  & 337.0  & 509.0  & 472.3  & /     & 423.3  \\
    \hline
    Usefulness & Exp.  & 5.0   & 4.7   & 4.7   & 4.7   & 5.0   & 4.7   & 5.0   & 5.0   & 1.0   & 5.0  \\
    \hline
    \end{tabular}}
    \vspace{-0.3cm}
  \label{table:user_study_results}%
\end{table}%

\noindent \textbf{Results.}
Table~\ref{table:user_study_results} reports the correctness and average completion time of each task for the experimental and control groups.
Compared with the control group, which has access only to compiler diagnostics, the experimental group completed tasks faster (271.7 vs. 416.4 seconds) and achieved higher correctness (21 vs. 14 correct answers).
We interview the control-group participants about how they locate corresponding implementations in the target codebase.
Two participants report that they first inspect compiler suggestions (if any), and then search the target codebase for identifiers with similar names or definitions.
When these simple attempts fail, they adopt the same strategy as \appname, i.e., finding pairs of code snippets involving the identifier and inferring the mapping.
The remaining participant uses this strategy directly.
This helps explain the shorter completion times and higher correctness of the experimental group: \appname automates a process that the control group performs manually and conducts large-scale searches to retrieve high-quality usage-function pairs for mapping inference.
For example, in T6, two non-existent identifiers need to be fixed, and locating the usage-function pairs is time-consuming because the corresponding functions in Neovim are renamed and placed in different files. 
Thus, the control group spends much longer than the experimental group.
By contrast, in T2, replacing \verb|STRLEN| (a custom macro in Vim) with \verb|strlen| is relatively straightforward, so the time gap between the two groups is small.

For tasks (i.e., T4/8/10) where \appname fails to provide the correct fix,
we find that the control group also struggles.
In T4/10, where \appname incorrectly reasons about the transformations demonstrated in the retrieved usage function pairs, developers also find these cases confusing.
For example, in T4, the control group makes the same mistake as \appname, i.e., they remove only \verb|in_vim9script()| but fail to remove the associated condition (see Figure~\ref{fig:example_failed_invim9script}).
In T8, \appname fails because no valid usage-function pairs demonstrating the mapping can be retrieved.
We observe that control-group participants abandon the task, stating that they cannot find relevant information on how to replace the non-existent identifier.

Regarding the usefulness of \appname, participants in the experimental group generally give ratings of four or five (except for T8, where \appname fails to retrieve any usage-function pairs).
Our interview suggests that these high ratings stem from \appname providing both the fix and verifiable usage-function pairs that explain the mapping, allowing developers to efficiently verify the generated fixes.

In addition, there are five cases (i.e., T1/2/4/6/9) where \ppathf exhibits other porting errors, such as patch-location mismatch or failure to port part of the patch logic.
We observe that all participants still port the patches correctly.
Participant feedback further indicates that porting errors other than implicit inconsistencies are usually easy to resolve, as they only require inspecting local context (e.g., the functions being patched).

\subsection{Lessons Learned}
\textbf{Global contextual knowledge across codebases is a fundamental challenge in patch porting, 
and effective automation should primarily focus on assisting developers in collecting such context.}
Our user study shows that developers can reliably and efficiently resolve porting errors confined to local context. 
In contrast, addressing implicit inconsistencies, 
which requires extablishing correct global mappings between semantically equivalent program elements across codebases,
consistently incur the highest time overhead.
Developers abandon tasks when no reliable mapping evidence can be identified.
\appname is effective precisely because it automates the collection of global contextual information by retrieving usage-function pairs that expose cross-codebase mappings.
However, \appname relies on the availability of such usage-based evidence and fails when no valid usage function pairs can be retrieved.
Future tools should support collecting richer forms of context (e.g., code evolution history) to further reduce the burden on developers in collecting the global context required for patch porting.

\noindent \textbf{Providing verifiable rationale for tool outputs is critical to the practical adoption of the tool.}
Our userstudy shows that developers require sufficient transparency to understand why a particular modification is applied before trusting and integrating it into production code.
Participants consistently emphasized the importance of being informed about the \appname’s underlying principles, as well as the evidence provided by \appname (e.g., the retrieved usage function pairs demonstrating how equivalent functionality is implemented across codebases).
Importantly, developers expect this evidence to be verifiable, rather than merely an explanation generated by the LLM.
Moreover, making intermediate results inspectable enables developers to efficiently follow the tool's logic and validate the correctness of the ported patch. 
Future patch porting tools should prioritize transparency and evidence-driven outputs over opaque automation, ensuring that developers can confidently participate in, audit, and ultimately trust the porting process.

\noindent \textbf{Controlling LLM hallucinations and limiting over-aggressive automation are essential for reliable patch porting.}
We observe that LLMs may exhibit overly proactive behaviors, i.e., introducing speculative or unintended code modifications, which can pose unexpected risks to the codebase and undermine developers’ trust in the tool.
In practice, developers prefer conservative repair behavior, where modifications made by the tool are precise, minimal, and well-justified.
Moreover, prompt inspection and feedback after each modification are crucial to preventing error accumulation and to avoid successive repairs diverging from the intended patch semantics.
To support such conservativeness, it is critical to incorporate validation mechanisms that constrain LLM hallucinations. 
In \appname, we employ a set of heuristics to validate each proposed modification, e.g., preventing the introduction of new non-existent identifiers, disallowing changes unrelated to the error location, and rejecting excessive or unnecessary edits.
A modification is accepted and allowed to proceed to the next iteration only when all validation conditions are satisfied.
Otherwise, \appname refrains from making changes and falls back to the original code.
Effective porting tools should adopt conservative modification strategies, relying on automatic validation and deferring to explicit human verification when necessary, rather than pursuing aggressive end-to-end automation.

\subsection{Threats to Validity}

\noindent \textbf{Internal validity.}
First, our benchmark is constructed by running existing patch porting tools and then manually retaining patches whose remaining errors are caused by implicit inconsistencies. This process may introduce selection bias and make the benchmark partially dependent on prior tool behavior and manual curation decisions.
Second, we manually assess fix correctness without behavioral validation through testing and treat developer-submitted patches as the ground truth. As a result, the assessment may be influenced by validator expertise, and any errors in the developer patches could further bias the validation results.

\noindent \textbf{External validity.}
Though our datasets cover two typical patch porting scenarios, only one project (pair) is involved in each scenario, i.e., the Vim-Neovim dataset for cross-fork porting and the Linux dataset for cross-branch porting.
Nevertheless, these two datasets are typical and exclusively investigated in existing works~\cite{pan2024automating, shariffdeen2021automated, yang2023enhancing}.
Future works are required to collect larger patch porting datasets, 
involving more projects written in different programming languages,
to more comprehensively evaluate the generalizability of \appname.

\section{Conclusion and Future Work} \label{sec:conclusion}
In this paper, we take the first step toward automating the mitigation of implicit inconsistencies in patch porting.
Our approach, \appname, enables collaboration among an LLM, a compiler, and code analysis utilities.
For inconsistencies caused by non-existent identifiers, \appname retrieves matched code-segment pairs involving the identifier as mapping knowledge to guide mitigation.
Experiments on cross-fork and cross-branch porting show that \appname effectively refines patches with unresolved implicit inconsistencies and fixes more than twice as many patches as the best-performing baseline.
The results also demonstrate its generalizability across different underlying LLMs.
We further conduct a user study with our industry partner to evaluate its practical value in patch porting workflows.
In future work, we plan to apply \appname to more projects and programming languages.

\section*{ACKNOWLEDGMENTS}

This research/project is supported by the National Natural Science Foundation of China (No.92582107) and Zhejiang Provincial Natural Science Foundation of China (No.LZ25F020003)

\balance
\bibliographystyle{ACM-Reference-Format}
\bibliography{reference}

\end{sloppypar}
\end{document}